\documentclass{aa}
\usepackage{graphicx}
\newcommand{\Tef}{T_{\rm eff}}
\newcommand{\EBV}{E_{\rm B-V}}

\begin{document}

\thesaurus{08(08.01.1; 08.16.4; 08.09.2: Sakurai's object)}

\title{
The spectrum of V4334 Sgr (Sakurai's object) in August, 1998
\thanks{Based  on  observations  collected at the European Southern
Observatory, La Silla, Chile}}

\author{Ya.V. Pavlenko\inst{1,2} \and H. W. Duerbeck\inst{3}}

\offprints{Ya.V. Pavlenko}

\institute{Main Astronomical Observatory of Academy of Sciences of
Ukraine, Golosiiv woods, Kyiv-127, 03680 Ukraine, e-mail: yp@mao.kiev.ua
\and
Isaac Newton Institute of Chile, Kiev Branch
\and
WE/OBSS, Free University Brussels (VUB),
Pleinlaan 2, B-1050 Brussels, Belgium, e-mail: hduerbec@vub.ac.be}

\date{Received ; accepted }

\authorrunning{Ya.V. Pavlenko and H.W. Duerbeck}
\titlerunning{The spectrum of V4334 Sgr in August, 1998}

\maketitle

\begin{abstract}
Theoretical spectral energy distributions are computed for a grid
of hydrogen-deficient and carbon-rich model atmospheres with a
range in $\Tef$ of $4000 - 5500$~K and $\log g$ of $1.0 - 0.0$,
using the technique of opacity sampling, and taking into account
continuous, atomic line and molecular band absorption. The
energy distributions are compared with the spectrum of V4334 Sgr
(Sakurai's object) of August 12, 1998 in the wavelength interval
$300-1000$ nm. This comparison yields an effective temperature of
V4334 Sgr of $\Tef = 5250\pm 200~\rm K$, and a value for the 
interstellar plus circumstellar reddening of $\EBV = 1.3\pm 0.1$.
\end{abstract}

\keywords{Stars: individual: V4334 Sgr (Sakurai's object) -- Stars:
AGB and post-AGB evolution -- Stars: model atmospheres -- Stars:
energy distributions -- Stars: effective temperatures --
Stars: gravities -- Stars: interstellar reddening}

\section{Introduction}
V4334 Sgr, the ``novalike object in Sagittarius'', was discovered by Y.~Sakurai
on February 20, 1996 (Nakano et al. 1996). Soon after discovery, it was
found to be a final He flash object, a star on the rare evolutionary track
that leads from the central stars of planetary nebulae back to the red giant
region, an object also called a ``born-again giant''. Its progenitor was a
faint blue star ($\sim$ 21$\rm ^m$) in the centre of a low surface brightness
planetary nebula (Duerbeck \& Benetti 1996). Early spectroscopic studies of
the object found an increasing hydrogen deficit in the atmosphere, and a
C/O ratio $>$ 1 (Asplund et al. 1997, Kipper \& Klochkova 1997).

\begin{figure}
\begin{center}
\includegraphics[width=73mm,angle=270]{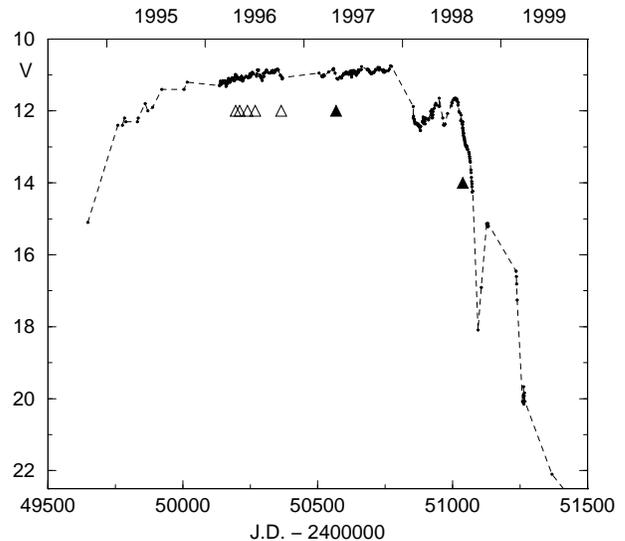}
\end{center}
\caption[]{The visual light curve of V4334 Sgr, according to Duerbeck et
al.~(2000). The times of five spectroscopic observations, analyzed by
Asplund et al.~(1997, 1999) and Kipper \& Klochkova (1997), are indicated by
open triangles. The times of our two spectroscopic observations of
April, 1997 and August, 1998 are marked by filled triangles. The second
observation took place at the beginning of a deep brightness decline.}
\end{figure}

Unlike the slowly evolving born-again giant FG Sge, which rose in
brightness during the first half of the 20th century, V4334 Sgr
is a very quickly evolving object (Fig.~1). It shows striking similarities
with the unfortunately poorly documented final He flash object
V605 Aql that erupted in 1918 -- 1923 (see Clayton \& de Marco (1997)
for an overview). Between March 1996 and April 1997, V4334 Sgr
evolved from early F to late red giant stages. Infrared observations by
Kimeswenger et al. (1997) and by Kamath \& Ashok (1999) show a noticeable
excess in the infrared K band starting March 1997, indicative of dust
formation. However, only in early 1998 a $\sim 1^{\rm m}$
depression appeared in the optical light curve (Liller et al. 1998,
Duerbeck et al. 1999, 2000).
This moderate depression persisted for a few months, and was followed by a
$\sim 6^{\rm m}$ decline which started in August 1998. After a partial
recovery in late 1998 -- early 1999, the dust obscuration continued with
increasing strength, and it may be assumed that in the year 2000, only the
short-wavelength tail of the cool dust envelope can be observed in
the optical spectrum, while the stellar photosphere is engulfed in
the opaque dust shell. Unfortunately, not much is know about
these late phases, since the spectroscopic coverage is poor due to
the extreme faintness of the object in the visible region.

A spectrum of V4334 Sgr was obtained on April 29, 1997, before any dust
obscuration in the optical region became evident, but after the strong IR
excess had evolved. This spectrum  was compared
with theoretical spectral energy distributions (Pavlenko, Yakovina
\& Duerbeck 2000). The authors derived a temperature of $\Tef =
5500~\rm K$, and an interstellar reddening of $\EBV = 0.7$. In this
paper, we apply the same methods to a spectrum which was taken 15 months
later.

\section{Observations}

On August 12, 1998, a spectrum (range $355 - 1000$~nm, resolving power
$\sim 2000$) was taken by A.~Piemonte with the B\&C Cassegrain
spectrograph attached to the ESO 1.52~m telescope, and kindly put at
the authors' disposal. It was taken around the time of the onset of the
$\sim 6^{\rm m}$ decline, at a time when V4334 Sgr was obviously already
suffering $\sim 1^{\rm m}$ of visual dust obscuration (Fig.~1).

The spectrum was reduced using standard IRAF routines, and the
spectral energy distribution (SED) was derived using spectrophotometric
standard stars. The observed SED is influenced by interstellar reddening as
well as by additional circumstellar reddening.
The value of the interstellar reddening is still not accurately known;
we will assume $\EBV = 0.7$ in the following, as was derived in the
analysis of the ``dustless'' spectrum of April 29, 1997, by
Pavlenko, Yakovina \& Duerbeck (2000); note, however, that
Duerbeck et al. (2000) prefer $\EBV=0.8$, a value which is also consistent
with the observed versus theoretical Balmer decrement of the planetary nebula
surrounding V4334 Sgr (Kerber et al. 2000).

We will assume that the dusty envelope acts like an additional
amount of interstellar reddening. Dereddened spectral energy
distributions of the August 12, 1998 spectrum were calculated
for assumed values $0.7 \le \EBV \le 1.3$, and compared with
model atmospheres.

\section{Procedure and Results}

Our computations of model atmospheres and synthetic spectra are carried
out in a classical approach: a plane-parallel model atmosphere in LTE,
with no energy divergence. We assumed chemical compositions of V4334 Sgr as
determined by Asplund et al. (1997) and Kipper \& Klochkova (1997).
Their results differ in some details, last not least because there 
are differences in the physical input parameters and in the details 
of the procedure.

\begin{table}
\caption{Abundances of H, He, C, N, O (scaled to $\sum N_i$ = 1)
used in this paper.}
{\small
\begin{tabular}{lllll}\hline\hline
Element  &   Asplund    &  Asplund  &  Kipper \&  & $\rm [C] = + 0.6$ \\
         &  et al.      & et al.    &  Klochkova  &                \\
         & (1997)       & (1999)    &  (1997)     & \\ \hline\hline
H & $-1.730$      & $-2.42$       &  $-2.10$  &   $-2.45$     \\
He& $-0.027$      & $-0.020$      &  $-0.01$  &   $-0.05$     \\
C & $-1.73$       & $-1.62$       &  $-2.01$  &   $-1.05$     \\
N & $-2.53$       & $-2.52$       &  $-2.70$  &   $-2.55$     \\
O & $-1.93$       & $-2.02$       &  $-2.59$  &   $-2.05$     \\
\hline\hline
\end{tabular}
}
\end{table}

The abundances  of {\it some} elements obtained by Asplund et al. (1997)
for May 1996 are given in the second column of
Table 1. Asplund et al. (1999) obtained another set of
abundances for October 1996. For comparison, we show the
abundances of Kipper \& Klochkova (1997), obtained independently
for a spectrum of July 1996.
In general, their results differ more from Asplund et al. (1997)
than the results of Asplund et al. (1997) and (1999).
Therefore, they provide a more interesting baseline for numerical
experiments carried out to clarify the dependence of our results
on input abundances.

From the common point of view, hydrogen is an important element,
and its abundance in the atmosphere of V4334 Sgr is steadily
declining (Asplund at al. 1997). However, Pavlenko et al. (2000)
showed that the dependence of our results on $\log N({\rm H})$ is
much weaker than that found by Asplund et al. (1997) for the early
stages of its evoluton. Indeed, in the case of lower  $\Tef~(\rm <
6500~K)$ the contribution of H and  H$^-$ to the total opacity
in the line forming region  is reduced in comparison with ``hot
models'', because molecular absorption which does not include
hydrogen compounds dominates there.

Furthermore, and more importantly, the input carbon
abundance of Asplund et al. (1997) is larger by up to ~0.6 dex
than that which was obtained by them from the fine analysis of high
resolution spectra. That is the well known ``carbon abundance
problem'' (Asplund et al. 2000), which has, unfortunately,
not yet been resolved.

One may think that, in the meantime, the model atmospheres should
{\em not} be computed with the derived C abundances by Asplund et
al. (1999) but rather with the higher input values. The use of the
derived lower C abundance would result in a new ``carbon problem''
and thus even lower C abundances would be derived. To clarify the
case, we studied the impact of higher carbon abundances on the
model atmospheres and SEDs of V4334 Sgr. We used 4
abundance sets (i.e. Asplund et al. (1997) for May 1996, Kipper
and Klochkova (1997) for July 1996, Asplund et al. (1999) for
October 1996, and the ``carbon rich case'' with [C] = + 0.6 with
respect to the Asplund et al. (1999) abundance) in order to study
different aspects of the problem:
\begin{itemize}
\item
to model the impact of abundance changes on our results,
\item
to avoid possible uncertainties related with the ``carbon problem''.
\end{itemize}

Opacity sampling model atmospheres and synthetical spectra of
V4334 Sgr were computed by the SAM941 (Pavlenko 1999) and WITA6
programmes, respectively. The model atmosphere computation
procedure is described in Pavlenko et al.~(2000) and Pavlenko
(2000). To clarify the topic, some details of the computations
procedure are given here:
\begin{itemize}
\item
Convection was treated in the frame of the mixing-length
theory with $l/H = 1.6$. Overshooting was not considered in the
computations;
\item
the list of atomic lines was taken from VALD (Piskunov et al.~1995);
\item
molecular opacities were computed in the frame of JOLA
approach. We take into account the absorption in the 20 band
system of diatomic molecules (see Table~2 in Pavlenko et al.~2000);
\item
bound-free absorption of C~{\sc i}, N~{\sc i} and O~{\sc i} atoms were
computed using cross-sections of Hoffsaess (1979);
\item
SAM941 and WITA6 use the same opacity lists.
\end{itemize}

Model atmospheres, synthetical spectra and SEDs were computed for a
microturbulent velocity of 5 km/s, which is a typical value for atmospheres
of post-AGB stars. The resulting theoretical spectra were convolved
by a gaussian with a half-width of 0.5 nm.

Basically, the atmospheric structure of chemically peculiar stars
should respond to abundance changes (Pavlenko \& Yakovina 1994,
2000). To study the impact of the set of input abundances on our
results, we carried out a few numerical experiments with model
atmospheres computed  with abundances of Asplund et al.~(1999) and
Kipper \& Klochkova (1997).

The comparison of the temperature structure of a model atmosphere
$\Tef/\log g = 5200/0.0$ computed for the four abundance grids is
shown in Fig.~2. We prefer to use pressure as a depth parameter, because
temperature structures in the coordinates ($\tau_{\rm ross}, T$)
look very similar (see Fig.~2b in Pavlenko et al.~2000). In
general, the photospheres are shifted outwards for higher carbon
abundances due to the opacity increase. Let us note a few items:
\begin{itemize}
\item
The structure of model atmospheres of V4334 Sgr, computed
for abundances by Asplund et al.~(1999), Kipper \& Klochkova
(1997), and the ``carbon-rich case'' differ substantially, especially
in the deep layers. Differences of model atmospheres computed for
Asplund et al.'s (1999) abundances obtained for May and October
of 1996 are rather marginal;
\item the structure of the outermost layers of model atmospheres
computed for the Asplund et al.~(1997) and Kipper \& Klochkova
abundances is similar. Moreover, the differences of the computed
SEDs are rather weak (Fig.~3). More pronounced differences are found
for the ``carbon-rich'' case;
\item in the blue region, the ``flux peaks'' which correspond to the
minima of molecular absorption, are less pronounced for the model
based on Kipper \& Klochkova's (1997) abundances, but even more so
in the observed spectrum (Fig.~3);
\item
We have to admit that there is an obvious lack of computed opacity
in the blue region, and the model is only correct in a qualitative sense
(Pavlenko \& Yakovina 2000).
\end{itemize}

\begin{figure}
\begin{center}
\includegraphics[width=88mm,height=140mm,angle=00]{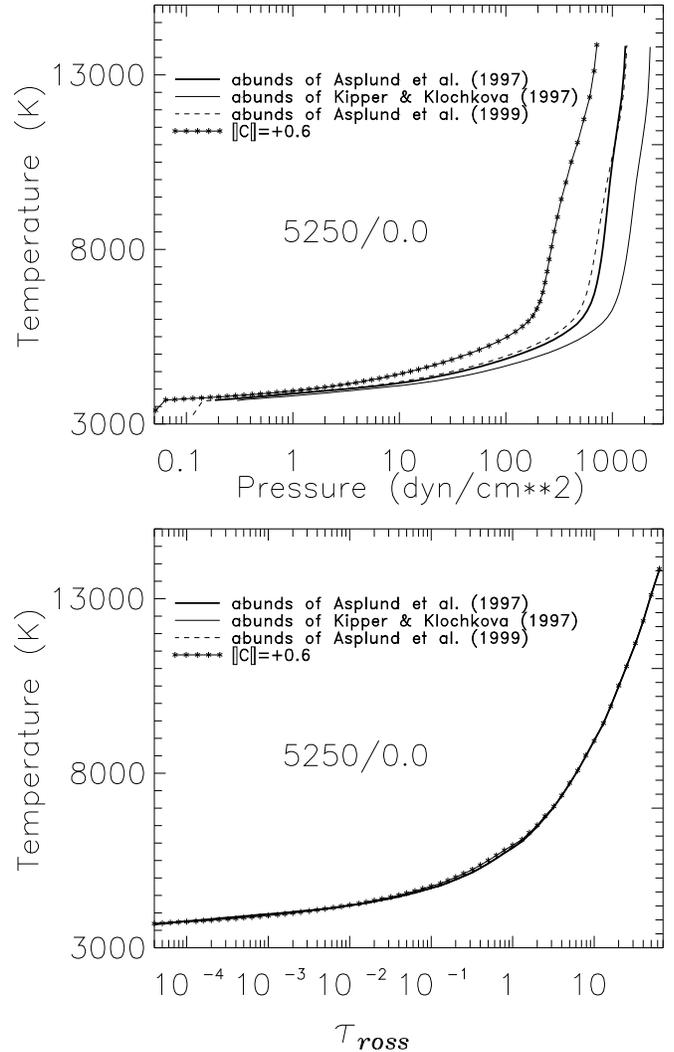}
\end{center}
\caption[]{The $T=f(P_g)$ structure of model atmospheres computed
for Asplund et al. (1997), Kipper \& Klochkova (1997), Asplund et
al. (1999), and [C] = + 0.6 abundances.  A second diagram
shows their $T=f(\tau)$ structure.}
\end{figure}

\begin{figure}
\begin{center}
\includegraphics[width=88mm,height=70mm,angle=00]{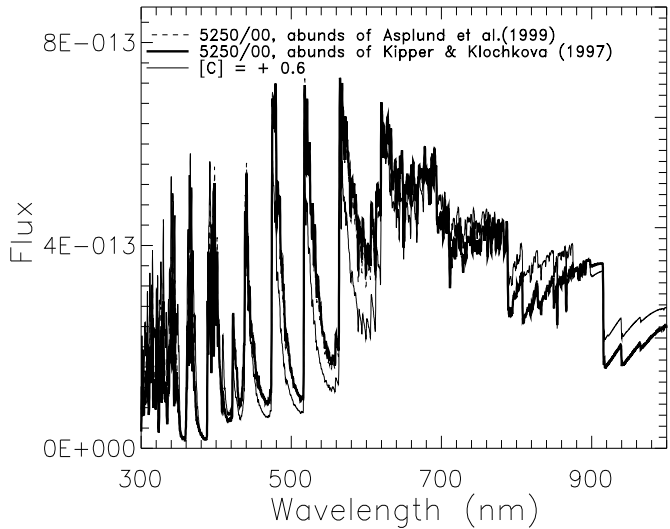}
\end{center}
\caption[]{SEDs computed for model atmospheres 5200/0.0 with
abundances taken from Asplund et al. (1999), Kipper \& Klochkova (1997)
and the ``carbon-rich'' case. The first two SEDs nearly coincide.}
\end{figure}

The overall shape of the spectrum of V4334 Sgr is governed by the
bands of the $\rm C_2$ Swan system and by the violet and red
systems of CN (Pavlenko et al.~2000). In the near UV ($\lambda <
400~\rm nm$), atomic absorption becomes important (see Pavlenko \&
Yakovina 2000).

Computed and observed spectra were normalized to equal flux
at $\lambda \sim$ 740 nm. In general, our main conclusions show a
rather weak dependence on the choice of $\lambda$ normalization
(see Pavlenko et al.~2000).

\begin{figure}
\begin{center}
\includegraphics[width=88mm,height=70mm,angle=00]{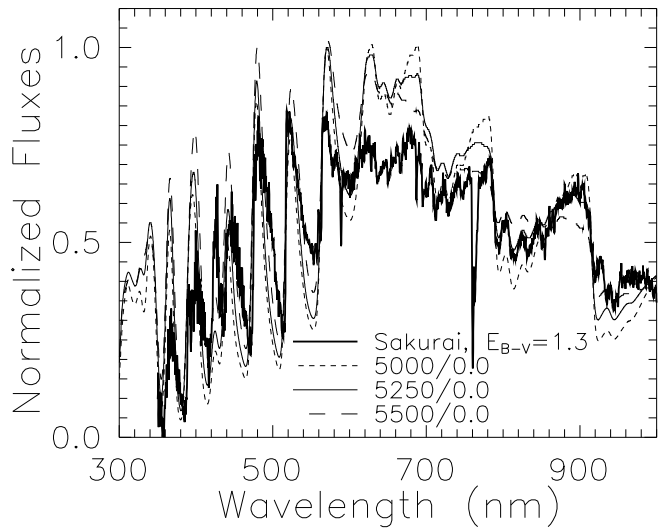}
\end{center}
\caption[]{
Synthetical spectra for different effective temperatures and $log g = 0$ 
are superimposed on the spectrum of V4334 Sgr (thick line) observed in 
August, 1998.}
\end{figure}

Nevertheless, the SED of V4334 Sgr depends critically on
$\Tef{}$ (Pavlenko \& Yakovina 2000). For the spectrum of August 12,
1998, a best fit was obtained for $\Tef \sim 5250~\rm K$ and $\EBV
= 1.3$ (Fig.~4). The fit is based on  the Asplund et al.
(1999) abundances for October 1996. The corresponding
values derived for our spectrum of April 29, 1997, are $\Tef =
5500~\rm K$ and $\EBV = 0.7$ (Pavlenko et al.~2000). In contrast
to the rapid cooling of V4334 Sgr in 1996, $\Tef$ has apparently
declined by only $\sim 250$~K in the course of the $15\frac{1}{2}$
months between early 1997 and mid-1998.

\section {Discussion}

Our SEDs were computed in the plane-parallel
approach. This may be used for $\Tef = 5500~\rm K$ (see discussions in
Asplund et al.~1998 and Pavlenko et al.~2000), but model atmospheres
for $\Tef = 5250~\rm K$ are more likely affected by sphericity effects.
Fits of model atmospheres  with $\log g = 0$ and $\log g =1$ to
the observed spectrum have a similar quality. This, however, only
indicates that the dependence of the model on gravity is rather weak.

It is noteworthy that our theoretical fluxes are much too
high between 600 and 700~nm, which suggests that an important
molecular line opacity is missing. A similar effect was noted by
Asplund et al.~(1997) in R CrB for $\lambda$ 500 nm.
It is doubtful that the assumption of
geometry and homogeneity is able to produce a shortage of flux at
such a very specific wavelength interval.
In V4334 Sgr, the effect is probably caused by
the absorption of one or more molecules which are not yet identified.

The observed and computed spectra for August,
1998 are more discrepant in the blue region, while the fit in the
red looks acceptable. A similar phenomenon was found in the investigation
of the April 1997 spectrum. By August 1998 the dust envelope had become thick,
while it was absent or marginal in April 1997. Thus we conclude that
the discrepancies are likely not caused by dust, but by one of the other
effects outlined above.
In the frame of our simple model, which assumes that the circumstellar
and the interstellar dust have similar properties in the visible part
of the spectrum, $E_{\rm B-V}$
increased from 0.7 to 1.3, according to our spectroscopic analysis.
Apparently, in 1997 there was no ``active''
optical circumstellar reddening in the line of sight, which is
proven by the similar effective temperatures derived from
spectroscopy and photometry (Fig.~2 of Asplund et al. 1997), while
at that time some dust clouds may have already formed outside the line of
sight, in order to account for the IR excess as observed by Kimeswenger et
al. (1997) and Kamath \& Ashok (1999). The increase in
$\EBV$ between April 1997 and August 1998, $\Delta \EBV = 0.6$, is
thus caused by circumstellar reddening of newly formed dust. If
the optical properties of this material are similar to those of
interstellar material, it should cause a drop in $V$ by $0.6
\times 3.1 \sim 1.85$ mag (with an uncertainty of $\pm 0.1$)
with respect to a ``dust-free'' lightcurve.

In order to estimate the magnitude of V4334 Sgr in such a
``dustfree'' lightcurve, the trend in the 1996-97 $V$ light curve
(as taken from Duerbeck et al. 2000) was linearly extrapolated to
August, 1998, and an ``expected'' magnitude of $V=10.75$ was
derived; if only the 1997 data are used for an extrapolation, the
result is $V=10.65$. In fact, the object is observed at $V=12.55$.
If we trust in the validity of the linear extrapolation, the
circumstellar $A_V$, which is necessary to put the object to the
observed brightness, is $1.85\pm 0.05$, in good agreement with the
value expected from the result of the spectroscopic study.

Is there another way to explain the observed spectrum of August, 1998,
e.g. by keeping the reddening at its original value of $\EBV =
0.7$, and decreasing the surface temperature? If $\Tef$ is
assumed to be as low as 4000~K, a general agreement with the overall SED
may be achieved, but the fit to the molecular bands is extremely
poor. The fit cannot be improved, even if the abundances of C, N,
O are permitted to vary (Fig.~5). Here we show
computations for different C/O ratios (given in logarithmic
scales). If O is more abundant than C, a completely
different molecular chemistry, and therefore
bands of oxygen compounds would appear in V4334 Sgr.
Fig.~5 illustrates that any changes of
carbon abundances do not result in a better fit of the
observed spectrum.

\begin{figure}
\begin{center}
\includegraphics[width=88mm,height=70mm,angle=00]{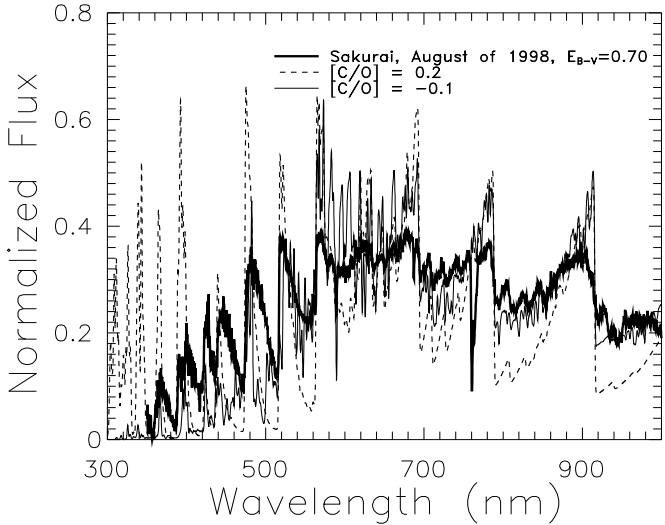}
\end{center}
\caption[]{Fits to SEDs with model 4000/0.0 of different
abundances. The expected very strong carbon bands can be
decreased in strength by reducing the C/O ratio, but the overall
fit would be still poor.}
\end{figure}

\begin{figure}
\begin{center}
\includegraphics[width=88mm,height=70mm,angle=00]{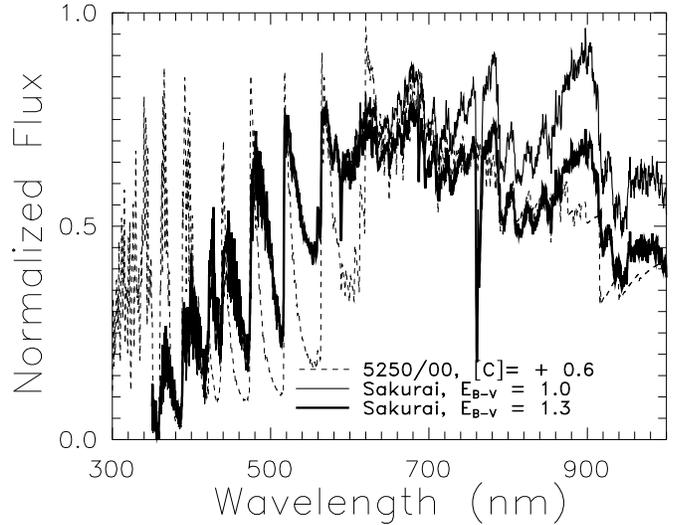}
\end{center}
\caption[]{Fits to SEDs of August 12, 1998 dereddened for $\EBV$ =
1.0 and 1.3 with model 5250/0.0 of the ``carbon-rich'' case. There
are some obvious problems of the fittings in the red and in the
blue.}
\end{figure}

\begin{figure}
\begin{center}
\includegraphics[width=72mm,angle=270]{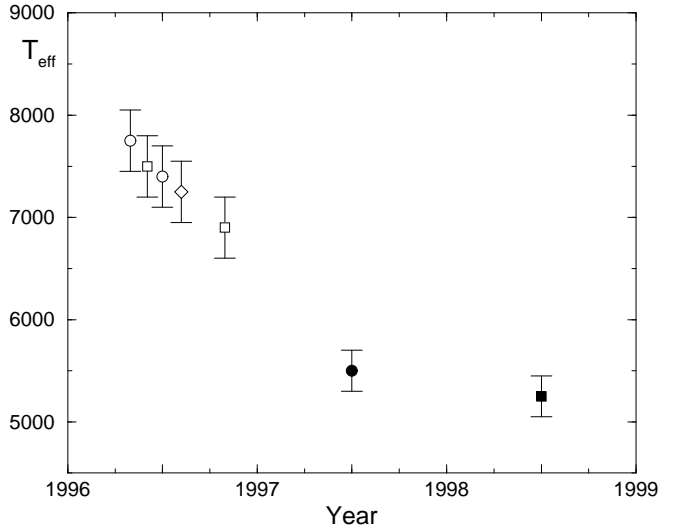}
\end{center}
\caption[]{Temporal changes of $\Tef$ in V4334 Sgr. Open circles:
Asplund et al. (1997), open squares: Asplund et al. (1999), open
diamond: Kipper and Klochkova (1997), filled circle: Pavlenko et al. (2000),
filled square: this paper. }
\end{figure}

The estimate of $\Tef{}$ depends on the assumed $\EBV$. Fig.~6
shows some fits of computed spectra to the observed spectrum of
V4334 Sgr, dereddened for $\EBV = 1.0$. There we also plot the SED of
V4334 Sgr, computed for the ``carbon-rich'' case.
This permits us to better assess the uncertainties in these
estimates. As we see in Fig.~6, our fit with $\EBV =1.0$ looks
less perfect both in the red and in the blue both for the Asplund
at al. (1999) abundances for May 1996 and even for the
``carbon-rich'' case. Neither variations of $\log N({\rm C})$ nor
of $\EBV$ permit to obtain a better fit to the SED observed in
August, 1998.

Thus we conclude that $\Tef$ of V4334 Sgr was near 5250~K in
August, 1998, and that the overall modification of the SED is caused by
circumstellar dust.

The temporal changes of $\Tef$ during 1996-1998 are shown in Fig.~7.
Following the assumption of Duerbeck et al.~(1998), that changes
of $\Tef$ were caused by the growth of the pseudophotosphere
surrounding a mass-losing object that radiates at approximately
constant luminosity, we conclude that its expansion had noticeably
slowed down or had even almost come to a halt during 1997 -- 1998,
and its temperature had almost
stabilized. This stage was accompanied by dust formation processes
in the envelope above the stellar photosphere, whose strength
increased with time. The object recovered only partly from these
events, so that by the second half of 1999 the photosphere of
V4334 Sgr had become basically invisible in the optical region.

\section{Conclusions}

A comparison of the observed spectra of V4334 Sgr in April, 1997
and August, 1998 with computed SEDs shows that the temperature of 
the radiating atmosphere had only decreased in $\Tef$ by 250~K, 
while the $\EBV$ had increased by 0.6.
This appears to be in perfect agreement with the observed
depression in the visual light curve, if a standard interstellar
extinction law $A_V = 3.1\times \EBV$ is assumed for newly formed
circumstellar dust. Thus there is
not only photometric, but also spectroscopic evidence that (a) the
light curve depressions are caused by dust, and that (b) the
luminosity and mass loss changed in such a way that the
pseudophotosphere was kept at an almost constant effective
temperature in 1997 --  1998.

\begin{acknowledgements}

We cordially thank Dr. A. Piemonte for taking the spectrum,
Prof. T. Kipper and Dr. M. Asplund for useful discussions,
and the VALD database team for helpful service. Partial
financial support of the investigations of Ya.~Pavlenko
was provided by a Small Research Grant of the American
Astronomical Society.

\end{acknowledgements}

\end{document}